\documentclass[aps,pre,twocolumn,groupedaddress,showpacs,floatfix]{revtex4}

\usepackage{psfig}

\begin{document}

\title{Phase synchronization and topological defects in inhomogeneous media}

\author{J\"orn Davidsen}
\email[]{joern.davidsen@utoronto.ca}
\author{Raymond Kapral}
\email[]{rkapral@chem.utoronto.ca}
\affiliation{Chemical Physics Theory Group, Department of
Chemistry, University of Toronto, Toronto, ON M5S 3H6, Canada}

\date{\today}

\begin{abstract}
The influence of topological defects on phase synchronization and
phase coherence in two-dimensional arrays of locally-coupled,
nonidentical, chaotic oscillators is investigated. The motion of
topological defects leads to a breakdown of phase synchronization
in the vicinities of the defects; however, the system is much
more phase coherent as long as the coupling between the
oscillators is strong enough to prohibit the continuous dynamical
creation and annihilation of defects. The generic occurrence of
topological defects in two and higher dimensions implies that the
concept of phase synchronization has to be modified for these
systems.
\end{abstract}


\maketitle

The rich collective behavior, including mutual entrainment and
self-synchronization, seen in systems of coupled oscillators has been a
stimulus for the long-standing interest in these systems (see, for example,
Refs.~\cite{pikovsky,strogatz00,kuramoto} and references therein). Recently,
attention has turned to the study of coupled \emph{chaotic}
oscillators and, in particular, to the phenomenon of phase
synchronization. Provided that the phase can be defined
\cite{pikovsky97,josic01}, two coupled nonidentical chaotic
oscillators are said to be phase synchronized if their frequencies
are locked \cite{rosenblum96,pikovsky}. This appears to be a
general phenomenon and it has been observed in such diverse
systems as electrically coupled neurons
\cite{elson98,makarenko98}, biomedical systems \cite{schaefer98},
chemical systems \cite{kiss02}, and spatially extended ecological
systems \cite{blasius99}, to name only a few. Moreover, the
potential role of phase synchronization in brain functions has
been recognized \cite{tass98,fitzgerald99}.

Much of the theoretical analyses of phase synchronization
have been carried out on systems consisting of two locally coupled
oscillators \cite{rosenblum96} or many globally coupled
oscillators \cite{pikovsky96}. Large one-dimensional chains
of locally coupled chaotic oscillators have been investigated very
recently \cite{liu01a,zhang01,valladares02}. Here, we address the
question whether phase synchronization can persist in higher
spatial dimensions where topological defects can play a
central role.  We show that the existence of topological defects
can lead to a breakdown of global phase synchronization in two-dimensional
arrays of nonidentical chaotic oscillators. While most
of the medium may remain phase synchronized, oscillators close
to moving topological defects have a different frequency. Despite
this fact, the phase coherence of the system is higher than in systems
without topological defects. The transition to phase synchronization
via phase clustering observed in one-dimensional systems \cite{zheng98,liu01a}
is not found in our simulations on two-dimensional systems; instead, a
transition involving point defects occurs.

Point topological defects in two-dimensional \emph{homogeneous}
oscillatory media are associated with the appearance of spiral
waves \cite{kapral,walgraef,mikhailov}. The phase field $\phi
({\bf r},t)$ of a medium with a spiral wave contains a point
topological phase defect in the spiral core such that
\begin{equation}
\frac{1}{2\pi }\oint \nabla \phi ({\bf r},t)\cdot d{\bf l}=\pm
n_{t},
\end{equation}
where $n_{t}$ is the topological charge of the defect
\cite{mermin79}. A topological defect corresponds to a point
in the medium where the local amplitude is zero and the phase is
not defined. For periodic boundary conditions, the net
topological charge of the medium is zero.
For identical chaotic oscillators \cite{textchaos}, spatially
coherent spiral dynamics can still exist implying a phase locking
of the oscillations \cite{goryachev00}.

To illustrate the phase synchronization properties in
two-dimensional networks of nonidentical chaotic oscillators, we
consider an $L \times L$ array of locally coupled R\"ossler
oscillators with periodic boundary conditions \cite{roessler76},
\begin{equation}
\frac{\partial {\bf x}({\bf r},t)}{\partial t} = {\bf R}({\bf
x}({\bf r},t)) + K \sum_{{\bf\hat{r}} \in N({\bf r})} \left ( {\bf
x}({\bf \hat{r}},t) - {\bf x}({\bf r},t) \right ),
\end{equation}
where $R_1 = -{\omega}({\bf r}) x_2 - x_3$, $R_2 = {\omega}({\bf
r}) x_1 + 0.2 x_2$, $R_3 = x_1 x_3 - 5.9 x_3 + 0.2$. The sites of
the lattice are labelled by ${\bf r}$, $K$ is the coupling
constant and $N({\bf r})$ is the set of the four next nearest
neighbors of site ${\bf r}$. The phase angles of the oscillators
are given by $\phi({\bf r},t) = \arctan \frac{x_2({\bf
r},t)}{x_1({\bf r},t)}$. To obtain the results presented below, we
take $L = 64$ and choose the ${\omega}({\bf r})$'s randomly from a
uniform distribution on the interval $[0.95,1.05]$, ensuring that
the system is in the chaotic regime.

\begin{figure}
\vspace{1.3cm} \psfig{figure=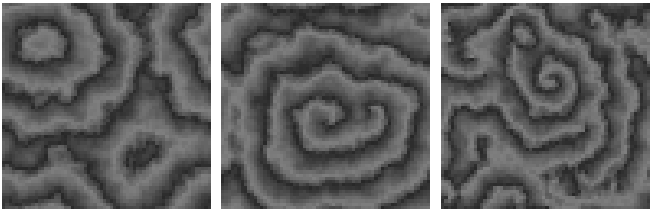,width=\columnwidth}
\caption{\label{figa} Snapshots of the phase field, all for the
same realization of $\omega({\bf r})$. From left to right: $K =
0.05$ without topological defects, $K = 0.05$ with two surviving
topological defects, and $K = 0.0419$. In the latter case,
topological defects are generated dynamically in pairs such that
the number of topological defects fluctuates around a value of 8
pairs.}
\end{figure}

First, we consider $K = K_0 \equiv 0.1$. For different initial
conditions leading to stable patterns, basically two scenarios can
be observed. For homogeneous initial conditions or initial
conditions with small inhomogeneities, no topological defects are
created and the system evolves to a target pattern similar to the
one in Fig. \ref{figa} (left panel). In the case of larger
inhomogeneities in the initial conditions, a number of topological
defects with $n_t = \pm 1$ is created initially. Some pairs of
topological defects with opposite charge are quickly annihilated
until only a small number is left generating a spiral pattern
similar to that in Fig.~\ref{figa} (middle panel). The topological
defects are not necessarily stationary and for very long times it
is possible that all defects could disappear through further
annihilation events; however, the motion of the surviving
topological defects is very slow compared to the initial
annihilation processes and no further annihilation events were
observed on our long simulation time scale ($10^4$ - $10^5$ spiral
revolutions). These two scenarios persist provided $K$ is not too
small.

\begin{figure}
\vspace{1.3cm} \psfig{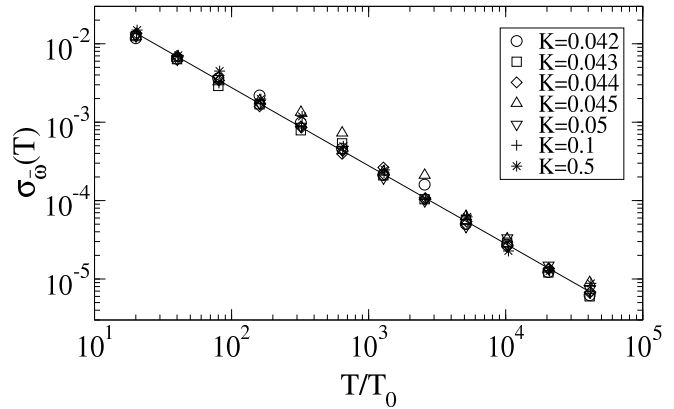}
\caption{\label{figb} Log-log plot of the standard deviation
$\sigma_{\bar{\omega}}(T)$ of the $\bar{\omega}(T)$ distribution
in the phase synchronized state describing the speed of
convergence towards $\omega_0$. The solid line with slope -1 is
plotted to guide the eye. Note that $T_0 = 2 \pi /\omega_0$
depends slightly on $K$.}
\end{figure}

When there are no topological defects, the system is phase
synchronized. The occurrence of phase synchronization for coupled
R\"ossler oscillators is usually attributed to the high degree of
phase coherence of the attractor of a \emph{single} R\"ossler
oscillator \cite{pikovsky} although this might not be a sufficient
condition \cite{josic01}. Phase coherence means that
$(\int^\infty_{-\infty} d\tau \langle \eta(t)
\eta(t+\tau)\rangle_t)^{1/2} \ll \omega_0$, where $d \tilde{\phi}/
d t = \omega_0 + \eta(t)$, $\tilde{\phi}$ is the unbounded phase
and $\langle \cdots \rangle_t$ signifies an average over $t$. If
$\eta$ is $\delta$-correlated Gaussian noise this condition
reduces to $\sigma \ll \omega_0$ and corresponds to a very narrow
peak of width $\sigma^2$ at $\omega_0$ in the power spectrum of
$x_1(t)$. However, generally temporal correlations in $\eta$
exist. These correlations determine the speed of convergence of
the time average $\bar{\omega}(T) = T^{-1}(\tilde{\phi} (T) -
\tilde{\phi} (0))$ towards $\omega_0$. For the system of coupled
R\"ossler oscillators considered here, the speed of convergence
--- as measured by the standard deviation of the ensemble
distribution of $\bar{\omega}(T)$ --- scales as $1/T$ as shown in
Fig.~\ref{figb}. This is in contrast to what one would expect if
the $\eta$ variables were independent where one obtains
$1/\sqrt{T}$ scaling since $\bar{\omega}(T) - \omega_0 = 1/T
\int_0^T dt \eta (t)$ and the standard deviation of $\int_0^T dt
\eta (t)$ scales with $\sqrt{T}$. The observed variation with $T$
is much faster, implying an extremely high degree of phase
coherence, even in the coupled system. The origin of this effect
resides in the shape of the local attractors. They are two-banded
with distinct frequencies $\omega_1$ and $\omega_2$ such that
$\omega_0 = (\omega_1 + \omega_2)/2$. Moreover, each deviation
from $\omega_1$ is followed by a deviation from $\omega_2$ of
similar amplitude but with opposite sign, and vice versa. Hence,
the deviations of $\bar{\omega}(T)$ from $\omega_0$ should be
dominated by the switching between the bands for not too large
values of $T$. Thus, the maximal deviation from $\omega_0$ is
given by the term $(\omega_1 - \omega_2)/(2 T)$, reproducing the
observed scaling. The two-banded structure can also be found in
the ensemble distribution of $\bar{\omega}(T)$. There, two
distinct peaks can be seen for certain $T$ values consistent with
the fluctuations around the $1/T$ scaling in Fig.~\ref{figb}.

\begin{figure}
\vspace{1.3cm} \psfig{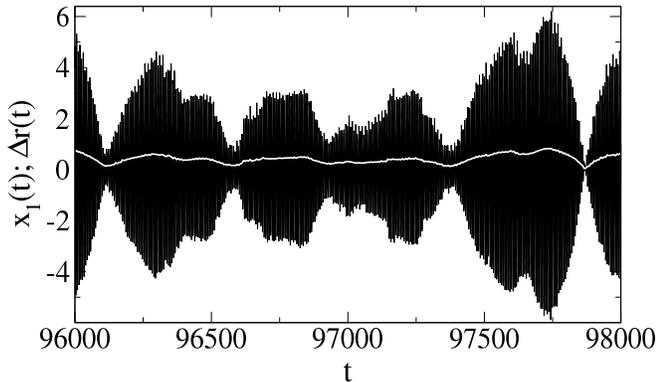}
\caption{\label{figc} Trajectory of $x_1(t)$ (black) for a fixed
oscillator and the distance $\Delta r(t)$ (white) of this
oscillator from the moving topological defect for $K=0.06$. The
two temporal behaviors are correlated, demonstrating the influence
of the topological defect's motion on the local orbit.}
\end{figure}

\begin{figure}
\vspace{1.3cm} \psfig{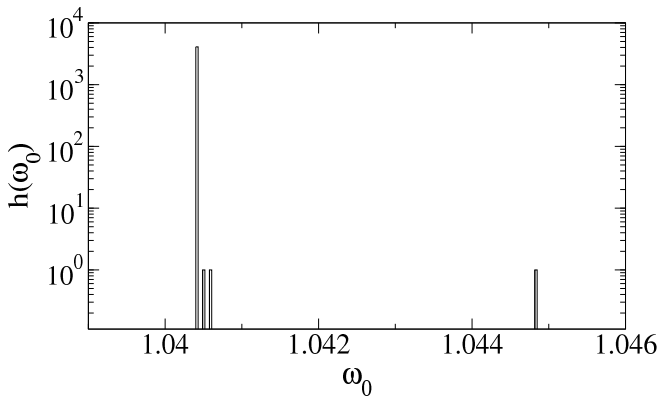}
\caption{\label{figd} The $\omega_0$ histogram $h(\omega_0)$ of
the oscillators for $K=0.06$ with two surviving topological
defects in the system. One defect moves causing the breakup of
global phase synchronization. The exact shape of the distribution
depends on the $\omega({\bf r})$ realization. Negative and/or
positive ``ouliners" can exist and correspond to oscillators in
the vicinity of the moving topological defect.}
\end{figure}

If topological defects are present global phase synchronization is
no longer guaranteed. Since a topological defect corresponds to a
point in the system where a phase variable cannot be defined, this
implies that the concept of phase synchronization can only be
applied to parts of the system which exclude topological defects.
For \emph{discrete} lattices of oscillators considered here, this
does not pose any problems because the defect is located almost
surely \emph{between} lattice sites. Thus, the presence of
topological defects does not exclude global phase synchronization
\emph{a priori}. We indeed observe phase synchronization for
quasi-stationary topological defects which explains the occurrence
of phase synchronization in the presence of topological defects
described in Ref. \cite{zhou02}. However, the motion of
topological defects destroys phase synchronization locally. As
shown in Fig.~\ref{figc}, the motion of a topological defect leads
to a distortion of nearby local orbits. This in turn influences
their local frequencies such that the system is no longer phase
synchronized as can be deduced from the corresponding $\omega_0$
distribution in Fig.~\ref{figd} \cite{text}. In particular, the
``outliners" in this distribution are located in the vicinity of
the topological defect(s).

For homogeneous oscillatory media it is well known that the time
scales for the motions of topological defects depend on the system
parameters, ranging from very slow motion to rapid dynamics
\cite{aranson02}. Inhomogeneity provides another mechanism for
defect motion and, in the present case of a network of
non-identical R\"ossler oscillators, this is the dominant
mechanism. For a network of identical R\"ossler oscillators, the
topological defects are quasi-stationary on the time scales
considered here; thus, the observed breakup of global phase
synchronization for this system is due to the quenched disorder in
the network.

\begin{figure}
\vspace{1.3cm} \psfig{figure=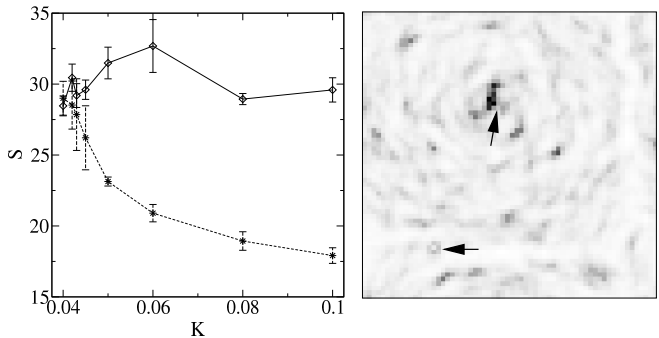,width=\columnwidth}
\caption{\label{fige} Left: $S$ as a function of $K$ for
homogeneous (star) and inhomogeneous (diamonds) initial
conditions. The results were obtained from an average over five
realizations of the random $\omega({\bf r})$ field. The phase
coherence is higher when (more) topological defects are present in
the system for $K > K_c \approx 0.043$. Below $K_c$ point defects
are created and annihilated dynamically. Right: $S({\bf r})$ for
the same $\omega ({\bf r})$ realization as in Figs.~\ref{figc} and
\ref{figd}. Dark areas correspond to high values of $S({\bf r})$.
The locations of the defects are given by the two arrows. The
lower left defect is the moving defect.}
\end{figure}

Although global phase synchronization is lost in the presence of
moving topological defects, the fluctuations of the instantaneous
period in the medium are greatly reduced. While the phase
coherence criterion, $(\int^\infty_{-\infty} d\tau \langle \eta(t)
\eta(t+\tau)\rangle_t)^{1/2} \ll \omega_0$, is satisfied to a high
degree of accuracy as in the case without topological defects, the
average amplitude of the fluctuations in the instantaneous periods
$T({\bf r},n) = 2 \pi / \omega({\bf r},n)$ as measured by $S
\equiv \langle\langle S({\bf r}) \rangle_{\bf r}\rangle_{\omega}$ with
\begin{equation}
S({\bf r}) = \frac{\langle T({\bf r},n) \rangle_n}{\sqrt{\langle
(T({\bf r},n) - \langle T({\bf r},n) \rangle_n )^2 \rangle_n}},
\label{eq2}
\end{equation}
are significantly different. Here, $T({\bf r},n)$ is the time
needed for the $n$th rotation at site ${\bf r}$ and $\langle
\cdots \rangle_{\omega}$ is the average over all realizations of
$\omega({\bf r})$. Figure~\ref{fige} (left panel) shows that the
value of $S$ is much larger above a certain coupling $K_c$ when
(more) topological defects are present in the system,
corresponding to a higher degree of coherence. An analysis of
$S({\bf r})$ shows that high values exist, especially in the
vicinity of quasi-stationary topological defects (see right panel
of Fig.~\ref{fige}). This is due to the influence of topological
defects on the shapes of the local orbits: they are close to
periodic limit cycles. This is in accord with observations in
\cite{goryachev96} for identical oscillators although the
``effective" spatial period doubling cascade found there is absent
in the present case.

The system maintains memory of its initial condition above $K_c
\approx 0.043$ leading to the different values of $S$.
For single $\omega({\bf r})$ realizations, different scenarios, with
either no defects or with defects present, can
be observed for $K_0 > K > K_c$ \cite{text2}. For $K$ below $K_c$,
however, a qualitatively new behavior occurs. Topological defects
are now generated and annihilated continuously by the dynamics
which is responsible for the loss of memory of the initial
condition. The number of topological defects fluctuates around a
mean value which increases with decreasing $K$. This is expected
because $K$ should control the characteristic length scale in the
system. Moreover, the topological defects move on average much
faster than above $K_c$, similar to defect mediated turbulence
\cite{coullet89}. The creation and annihilation of topological
defects as well as the fast motion lead to strong distortion of
the local orbits affected by the moving defects . This makes it
impossible to define a proper phase variable for most of the
oscillators in the network. Consequently, the concept of phase
synchronization can no longer be applied.

In Refs.~\cite{zheng98,liu01a}, the transition to phase
synchronization via the route of phase clustering and merging of
clusters was investigated for chains of oscillators. Liu, et al.
\cite{liu01a} concluded that phase clustering should be more
prevalent than full phase synchronization, especially in networks
of coupled neurons. However, the two-dimensional system considered
here shows a different behavior for increasing $K$ starting from a
value below $K_c$. We observe a transition from a state where
topological defects are generated continuously by the dynamics to
a partial phase synchronized state with moving topological
defects, similar to the case with surviving point defects. This
partial phase synchronized state is characterized by a large
number of phase synchronized oscillators and a small number of
oscillators close to the point defects with mutually different
frequencies. Thus, only one cluster with more than one oscillator
exists in contrast to the chain geometry. The existence of
topological defects in two dimensions and higher dimensions
suggests that the transition described here should be prevalent
for general networks of chaotic oscillators.

We have shown that the occurrence of topological defects in
two-dimensional arrays of nonidentical chaotic oscillators has a
two-fold effect: moving topological defects lead to a breakup of
global phase synchronization, while all topological defects
--- especially quasi-stationary ones --- increase the phase
coherence of the system significantly. Since moving and
quasi-stationary topological defects are present in many
circumstances, the apparently contradictory scenario of increased
phase coherence but loss of global phase synchronization occurs.

\begin{acknowledgments}
This work was supported in part by a grant from the
Natural Sciences and Engineering Research Council of Canada.
We thank G. Rousseau for providing the numerical integrator.
\end{acknowledgments}


\begin{thebibliography}{29}
\bibitem{textchaos}
The motion of chaotic oscillators cannot be uniquely
parametrized by the phase in contrast to classical oscillators;
however, the phase variable suffices for the determination of the
location and charge of a topological defect in the
spatially-distributed medium.
\bibitem{text}
The distortion of the local orbits by the motion of the
topological defects cannot only cause the breakdown of global
phase synchronization but also the breakdown of a proper phase
description, i.e., no unique center of rotation exists. However,
the percentage of loops in phase space where $(0,0)$ lies outside
the loop is of the order of $0.1\%$ for $K$ close to $K_0$ and,
thus, extremely small. In particular, the effect on the estimate
of the oscillator's frequency is negligible.
\end{thebibliography}
\end{document}